\documentclass[a4paper,11pt]{article}
\usepackage{graphicx}
\usepackage{amsmath}
\usepackage{amssymb}
\usepackage{amsfonts}

\title{Formation of vortices in a dense Bose-Einstein condensate}
\author{V.P. Barros $^\dagger$, Ch. Moseley $^{\dagger \dagger}$,
A. Gammal $^\dagger$, K. Ziegler $^{\dagger \dagger}$}

\begin{document}
\maketitle

\begin{quote}
 {\it $\dagger$Instituto de F\'isica, Universidade de S\~ao Paulo C.P. 66318, 05315-970 S\~ao 
Paulo, Brazil \\
$\dagger \dagger$ Institut f\"ur Physik, Universit\"at Ausgburg, D-86135 Augsburg, Germany}
\end{quote}

\begin{abstract}
A relaxation method is employed to study a rotating dense Bose-Einstein condensate beyond 
Thomas-Fermi approximation. We use a slave-boson model to describe the strongly interacting
condensate and derive a generalized non-linear Schr\"odinger equation with kinetic term for the rotating
condensate. In comparison with previous calculations, based on Thomas-Fermi approximation, significant
improvements are found in regions, where the condensate in a trap potential is not smooth. 
The critical angular velocity of the vortex formation is higher than in the Thomas-Fermi prediction.
\end{abstract}


\section{Introduction}

The appearance of vortices in a rotating dilute Bose-Einstein condensate (BEC) is a 
phenomenon that is intrinsically related to the existence of superfluidity 
\cite{Feder1,Feder2}. This has been intensively studied \cite{Dum,Mateus} after the achievement 
of BEC in alkali-metal gases \cite{Anderson,Bradley}. 
The vortex formation can be understood as a local suppression of the BEC due to rotation. 
It happens when the angular velocity $\Omega$ exceeds a critical value 
$\Omega_{c}$ \cite{Fetter}. The critical angular velocity is obtained by an energy argument
\cite{Fetter}, in which the energy of the BEC without a vortex is higher than the
energy of the BEC with a vortex for $\Omega>\Omega_{c}$. With higher angular
velocities more than one vortex can be created to minimize the energy of the BEC.
In this paper we will concentrate on values of $\Omega$ where only the
the creation of a single vortex is possible.   

Since vortices are accompanied by a local suppression of the
BEC, other mechanisms of BEC suppression may interfere with the vortex formation. An important
example is the depletion of the BEC by a strong interaction between the bosonic
particles. The latter is possible in the form of repulsive
s-wave scattering in a dense BEC \cite{shukla1}.

In the case of a dilute BEC (i.e. for weak interaction) a mean-field theory in terms of
the Gross-Pitaevskii equation provides a description of the vortex formation \cite{Dalfovo1}.
The relation between the number of particles in the BEC and the critical angular
velocity was determined within this approach \cite{Dalfovo2}.   
This can be extended to a dense Bose gas, where not all particles contribute to
the BEC (i.e. there is a depletion of the BEC). The discussion of the dense Bose gas
requires a statistical description of the entire Bose gas, using, for instance, a 
functional-integral representation \cite{ziegler0}. Nevertheless, it turns out that a classical
field equation for the BEC order parameter can also be derived, which is equivalent
with the Gross-Pitaevskii equation in the limit of a dilute Bose gas \cite{shukla1}
and a renormalized Gross-Pitaevskii equation for higher densities. In the regime of high 
densities the strongly interacting Bose gas had been studied within this statistical approach 
\cite{ziegler0,shukla1,ziegler1,ziegler2,stoof03,SBmanuscript,review}. 
A relatively simple method
to study strong interaction is the slave-boson approach. This method was originally suggested
for strongly interacting fermions \cite{barnes76,kotliar86} and was later also applied to
bosons \cite{ziegler0}. The central idea of the method is to construct operators or fields for
particles and holes in a fictitious vacuum, and to introduce a constraint, which guarantees
that each position is occupied either by a particle or by a hole. This leads to a dynamics,
where tunneling of physical particles is described by an exchange of a particle with a hole,
and no multiple occupation by particles is possible.
Although this construction makes the theory more complex (because of the additional holes),
it also provides more freedom to choose a proper mean-field approximation. Moreover, the bosonic
version of the slave-boson approach, which will be applied in this paper, allows for an
exact treatment of the constraint \cite{ziegler0}. Here we will not present the details of 
the slave-boson approach but refer to the literature. A recent discussion can be found in
Ref. \cite{SBmanuscript}. 
Within a mean-field
(Thomas-Fermi) approximation of a slave-boson theory we were able to study the formation 
of a vortex in the BEC of a dense Bose gas \cite{SBmanuscript}. However, a mean-field approach 
is a reliable approximation only for a BEC that is smoothly changing in space. The latter
is not the case near the core of a vortex, where the condensate density is changing
strongly on short scales. Therefore, it is important to extend the mean-field theory to
a classical field theory with a kinetic term. This leads to a nonlinear partial 
differential equation which can be treated with a numerical method. 
In the present work we are using a relaxation approach.
Relaxation techniques have been extensively used to obtain the ground-state solution of the
Gross-Pitaevskii equation \cite{Ruprecht,Gammal,Band,Adhikari,Chiofalo}. These techniques 
require interaction and need little computational efforts. The method provides stable 
solutions of the slave-boson approach. 

This work is organized as follows. In Section 2  
we summarize the main equation of slave-boson approach. The trapped condensate is 
shown in spherical and cylindrical coordinates in Section 3. In Section 4 we discuss the 
numerical procedure used in this work and present the numerical results with and 
without vortex. Afterwards we review our main results in section 5.

\section{Slave-Boson Approach}

We consider a grand-canonical ensemble of bosons with chemical potential $\mu$ 
at temperature $T=1/k_B\beta$.
Its thermodynamics is described by the partition function
\begin{equation}
Z=Tr e^{-\beta (H-\mu N)} \; ,
\end{equation}
where $H$ is the Hamiltonian of the interacting Bose gas, $N$ the particle number operator, 
and $Tr$ is the trace over all possible quantum states of the system.
If the interaction is due to a hard core of a given radius $a$, the Bose gas can be approximated by
a lattice gas with lattice constant $a$. Then we can apply the slave-boson approach 
\cite{ziegler0,shukla1,ziegler1,ziegler2,stoof03,SBmanuscript,review} which allows 
us to write the partition function $Z$ as a functional integral with respect to a complex (order-parameter)
field $\Phi({\bf r})$. Here we start with an expression for $Z$,
whose derivation was given in Ref. \cite{SBmanuscript}:
\begin{equation}
Z=\int e^{-S}\prod_r d\Phi({\bf r})
\label{partition0}
\end{equation}
with the continuum action
\begin{equation}
 S = \int \left\{ \beta' \Phi^\ast({\bf r}) \left[
 -\frac{\alpha}{(1+\alpha)^2} \frac{a^2}{6} \nabla^2 + \frac{1}{1+1/\alpha}
 \right] \Phi({\bf r}) - \log Z({\bf r}) \right\} \, {\rm d}^d r \; ,
 \label{Scontinuous}
\end{equation}
where
\begin{equation}
 Z({\bf r}) = \int e^{-\beta' \varphi({\bf r})^2} \, \frac{
 \sinh \left[\beta'\sqrt{(\varphi({\bf r}) + \mu'/2)^2+|\Phi({\bf r})|^2}\right]}{\left[ \beta'
 \sqrt{(\varphi({\bf r}) + \mu'/2)^2+|\Phi({\bf r})|^2}\right]} \, {\rm d}\varphi({\bf r}) \; .
 \label{function_Z}
\end{equation}
$\alpha$ is a free parameter which appears as an ambiguity in the functional integral \cite{note1}.
It is necessary for the integration,  
but the exact functional integral $Z$ does not depend on the choice of $\alpha$.
However,  approximations may depend on it. In agreement with a previous work \cite{shukla1} 
we choose $\alpha=1/5.5$ for the following calculation.
In $S$ we have introduced the parameters
\begin{equation}
 \beta' = \alpha J\beta \; , \quad \mu' = \frac\mu{\alpha J} \; ,
\end{equation}
where $J$ is the tunneling rate of the lattice model (i. e. before taking the continuum limit). 

Since $\Phi$ is the order-parameter field, the number of condensed bosons can be written as
an integral over $|\Phi({\bf r})|^2$:
\begin{equation}
 N_0 = \frac 1{a^3}\, \frac 1{(1+1/\alpha)^2} \int |\Phi({\bf r})|^2 {\rm d}^3r \; .
\label{number}
\end{equation}
Thus we can associate $\frac{|\Phi({\bf r})|^2}{a^3(1+1/\alpha)^2}$ 
with the density of condensed bosons.

The evaluation of the partition function in Eq. (\ref{partition0}) requires a functional
integration. This can be performed approximately by a saddle-point integration. The
saddle point is determined as the solution of a vanishing variation of S (i.e. $\delta S=0$).
This leads to
\begin{equation}
 \left[ - \frac{Ja^2}{6} \nabla^2 + (1+\alpha)J - \alpha J \frac{(1+1/\alpha)^2}{\beta'} \,
 \frac { \partial  \log Z({\bf r})} {\partial (|\Phi({\bf r})|^2)} \right] 
\Phi({\bf r}) = 0 \; ,
 \label{SBoriginal}
\end{equation}
where the last term can be explicitly evaluated as
\begin{equation}\label{4}
\chi \equiv\frac{\partial \log Z(\bf r)}{\partial |{\Phi}({\bf 
r})|^2}=
\frac12 \frac{1}{Z(\bf r)}
\int_{-\infty}^{+\infty}d\varphi({\bf r}) e^{-\varphi({\bf r})^2}
\left[ \frac{\cosh \gamma}{\gamma^2}-\frac{\sinh \gamma}{\gamma^3} \right], 
\end{equation}
with
\begin{equation}\label{5}
\gamma=\sqrt{ \bigl(\varphi({\bf r})+ \mu'/2\bigr)^2+|{\Phi}({\bf r})|^2 }.
\end{equation} 
Eq. (\ref{SBoriginal}) corresponds to the Gross-Pitaevskii equation 
\cite{pitaevskii}
\begin{equation}
 \left[-\frac{\hbar^2}{2M} \nabla^2 - {\mu_{GP}} + V({\bf r}) +
 g|\Phi({\bf r})|^2 \right] \Phi({\bf r}) = 0 \; ,
 \label{conventionalGP}
\end{equation}
The latter is obtained
as a special case of Eq. (\ref{SBoriginal}) for a weakly interacting Bose gas \cite{SBmanuscript}.
Here $M$ is the mass of the bosons, and the interaction strength of the two-particle
interaction $g$ is related to the $s$-wave scattering length $a_s$ by
\begin{equation}
 g = \frac{4\pi a_s \hbar^2}M \; .
\end{equation}
There is a direct correspondence between the parameters $J$, $\mu$, and $a$ in Eq.
(\ref{SBoriginal}), and the parameters $M$, ${\tilde\mu}$, and $g$ in Eq. (\ref{conventionalGP})
\cite{review}
\begin{equation}
 \frac{\hbar^2}{2M} \equiv \frac{Ja^2}6 \ ,
\hskip0.5cm
 g \equiv 2a^3J \ ,
 \hskip0.5cm
 {\mu_{GP}} \equiv \mu - J \; .
\label{conection}
\end{equation}
Since Eq. (\ref{SBoriginal}) describes the condensate for arbitrary density, we will use this
in the subsequent calculation to evaluate the condensate profile of a trapped and rotating dense
Bose gas.

\section{Dense condensate in a spherical harmonic trap potential}

We assume a spherical trap potential given by
\begin{equation}
 V({\bf r})=\frac M2 \omega_{\rm ho}^2 {\bf r}^2 \; ,
 \label{trap}
\end{equation}
where $\omega_{\rm ho}$ is the trap angular frequency
measured in rad/s. 
In typical experiments, the oscillator length $d_{\rm ho}=\sqrt{\hbar/M\omega_{\rm ho}}$ is of the
order of a few $\rm \mu m$ \cite{Bpitaevskii}. 
The external potential can now be included in a space-dependent chemical potential as
$\mu\rightarrow\mu({\bf r})=\mu-V({\bf r})$.

Considering, for instance, $^{85}$Rb atoms near a Feshbach resonance 
\cite{wieman3}, we can study a Bose gas in a dense regime with a scattering length 
$a_s\sim a\sim 200\rm nm$. In our calculations we choose the parameters
\begin{equation}
 \beta'=1 \; , \quad \frac{k_{\rm B}T}{\hbar\omega_{\rm ho}} = 36.93 \; , \quad
 \frac a{d_{\rm ho}} = 0.1215 \; 
 \label{parameters}
\end{equation}
in order to compare with previous results in Ref. \cite{SBmanuscript}.

In a spherical symmetric trap we can assume that the condensate is also spherically
symmetric:
\begin{equation}
 \Phi({\bf r}) = \phi(r) \; .
\end{equation}
Then Eq. (\ref{SBoriginal}) takes the form
\begin{equation}
 \left[-\frac{a^2}{6\alpha} \left(\frac{\partial^2}{\partial r^2}
 + \frac 2r \, \frac{\partial}{\partial r} \right) + (1+1/\alpha)
 - \frac{(1+1/\alpha)^2}{\beta'} \, \frac\partial{\partial\left(|\phi(r)|^2\right)} 
 \log Z(r)\right]\phi(r) = 0 \; .
 \label{sbeq_novortex}
\end{equation}
Now we consider a Bose gas which rotates with angular velocity $\Omega$ about the $z$-axis.
In the rotating frame this can be described by adding
$-\hbar\Omega \hat{L}_z$ to the kinetic energy in Eq. (\ref{SBoriginal}), where the 
$z$-component of the angular momentum operator is
\begin{equation}
 \hat{L}_z = -{\rm i}\left( x\frac\partial{\partial y}-y\frac\partial{\partial x} \right)
 = -{\rm i} \frac\partial{\partial\varphi} \; ,
\end{equation}
and $\varphi$ is the polar angle in cylindrical coordinates.

For a condensate with a single vortex along the $z$-axis, we use cylindrical
coordinates $(r_\perp,z,\varphi)$ and make the ansatz
\begin{equation}
 \Phi({\bf r}) = \phi(r_\perp,z) e^{{\rm i } m \varphi}
\end{equation}
for the condensate order parameter. For $m=0$ there is no vortex and for $m=1$ there is a single vortex. 
Then the classical field equation reads
\[
 \bigg[ -\frac{a^2}{6\alpha} \left(\frac{\partial^2}{\partial r_\perp^2} + \frac 1{r_\perp} \, 
 \frac{\partial}{\partial r_\perp} + \frac{\partial^2}{\partial z^2} \right) + (1+1/\alpha)
 + \left(\frac{a^2 m^2}{6\alpha r_\perp^2} - m\Omega'\right)
\]
\begin{equation}
  - \frac{(1+1/\alpha)^2}{\beta'} \, \frac\partial{\partial\left(|\phi(r_\perp,z)|^2\right)} 
 \log Z(r_\perp,z)\bigg] \phi(r_\perp,z) = 0 \; ,
 \label{sbeq_vortex}
\end{equation}
where $\Omega'$ is the rescaled angular velocity
\begin{equation}
 \Omega' \equiv \frac{\hbar\Omega}{\alpha J} \; .
\end{equation}
The action is expressed as
\begin{eqnarray}
 S(\Omega')&=&2\pi\int \biggl\{
\beta'\biggl[-\frac{\alpha}{(1+\alpha)^2}\frac{a^2}{6}\phi^\ast\nabla^{2}_{r_{\perp},z}\phi+\biggl(
\frac{\alpha}{(1+\alpha)^2}\frac{a^2 m^2}{6 r_\perp ^2}-m\Omega'\biggr)|\phi|^{2} +
\frac{|\phi|^{2}}{(1+\alpha)^2} \biggr]\nonumber\\ & &-\log Z(r_\perp,z) \biggr\} dr_\perp dz
\end{eqnarray}

\section{Numerical results}

The slave-boson equations (\ref{sbeq_novortex}) and (\ref{sbeq_vortex}) for a trapped
condensate without and with vortex, respectively, can be solved numerically by means of a 
relaxation algorithm. The full solutions are compared to previous results,
which were calculated from the Thomas-Fermi approximation (TFA) \cite{SBmanuscript}. In 
the TFA it is assumed that the kinetic energy can be neglected in comparison to the
potential energy \cite{Bpitaevskii}. Thus the kinetic term containing the spatial derivatives is
neglected. This leads to a transcendental equation which is easier to solve than the full
differential equation.

The relaxation algorithm which we employ now is similar to 
that applied for the GP equation in Ref. \cite{marijana}, scheme B.
First, we introduce an artificial dynamics by writing Eq. (\ref{sbeq_vortex}) in a time-dependent 
form as
\[
\frac{\partial {\tilde \phi}(r_{\perp},z)}{\partial t}=
\] 
\begin{equation}
\left[-\frac{a^2}{6\alpha}\nabla^{2}_{r_{\perp},z} + (1+1/\alpha)
+\left(\frac{a^2 m^2}{6\alpha r_\perp^2} - m\Omega'\right)
- \frac{(1+1/\alpha)^2}{\beta'} 
 \chi\right]{\tilde \phi}(r_{\perp},z) \; .
 \label{SBtime}
\end{equation}
where $|\tilde{\phi}|^{2}\equiv|\phi|^{2}/\eta$, 
$\eta\equiv\int|\phi|^2d^3r$. With this substitution the  
function $\tilde\phi$
will always be
normalized to 1 and bounded to values not too big or too small to be 
operated numerically by the kinetic term. This provides more stability to 
the code.   

Eq. (\ref{SBtime}) is discretized in the split-step form
\begin{eqnarray}
\left.
\begin{array}{ll}
\tilde{\phi}_{n+1/3}\leftarrow\tilde{\phi}_n+\frac{3\alpha\Delta t}{a^{2}}\biggl[\biggl(\frac{\alpha+1}{\alpha}\biggr)
+\left(\frac{a^2 m^2}{6\alpha r_\perp^2} - m\Omega'\right)
-\biggl(\frac{\alpha +1}{\alpha\beta'^{2}}\biggr)^2 \chi_n \biggr] \\
\tilde{\phi}_{n+2/3}\leftarrow {\cal{O}}_{CN} \tilde{\phi}_{n+1/3}\\
\tilde{\phi}_{n+1}\leftarrow\tilde{\phi}_n
+\frac{3\alpha\Delta t}{a^{2}}\biggl[\biggl(\frac{\alpha+1}{\alpha}\biggr)+\left(\frac{a^2 m^2}{6\alpha r_\perp^2} 
- m\Omega'\right)-\biggl(\frac{\alpha +1}{\alpha\beta'^{2}}\biggr)^2 \chi_n \biggr]\\
\eta_{n+1}\leftarrow \eta_{n}\int |{\tilde \phi}_{n+1}|^{2}d^{3}r\\
{\tilde \phi}_{n+1}\leftarrow \frac{{\tilde \phi}_{n+1}}{\int |{\tilde 
\phi}_{n+1}|^{2}d^{3}r}
\end{array}
\right\},
\end{eqnarray}
where the subscripts containing $n$ correspond to the $n^{th}$ time step.  ${\cal{O}}_{CN}$ corresponds to the 
Crank-Nicolson evolution algorithm alternated in $r_{\perp}$ and $z$ directions \cite{koonin}. In the 
$r_{\perp}$ direction, Neumann boundary conditions ($\frac{\partial \phi (r_{\perp}=0,z)}{\partial 
r_{\perp}}=0$) were taken at the origin. The integrals of $\chi$ and $Z$ can be evaluated efficiently at each 
time step by Gauss-Hermite quadrature with good convergence using only nine points. The algorithm starts with 
an initial ansatz distribution loaded to $\tilde{\phi}$. After evolving for sufficient long time the time 
derivative Eq. (\ref{SBtime}) vanishes, $\tilde{\phi}_{n}$ converges to the stationary solution and $\eta_n$ 
converge to the norm of ${\phi}$.

\begin{figure}[ht]
\centering
\scalebox{0.5}{
\includegraphics{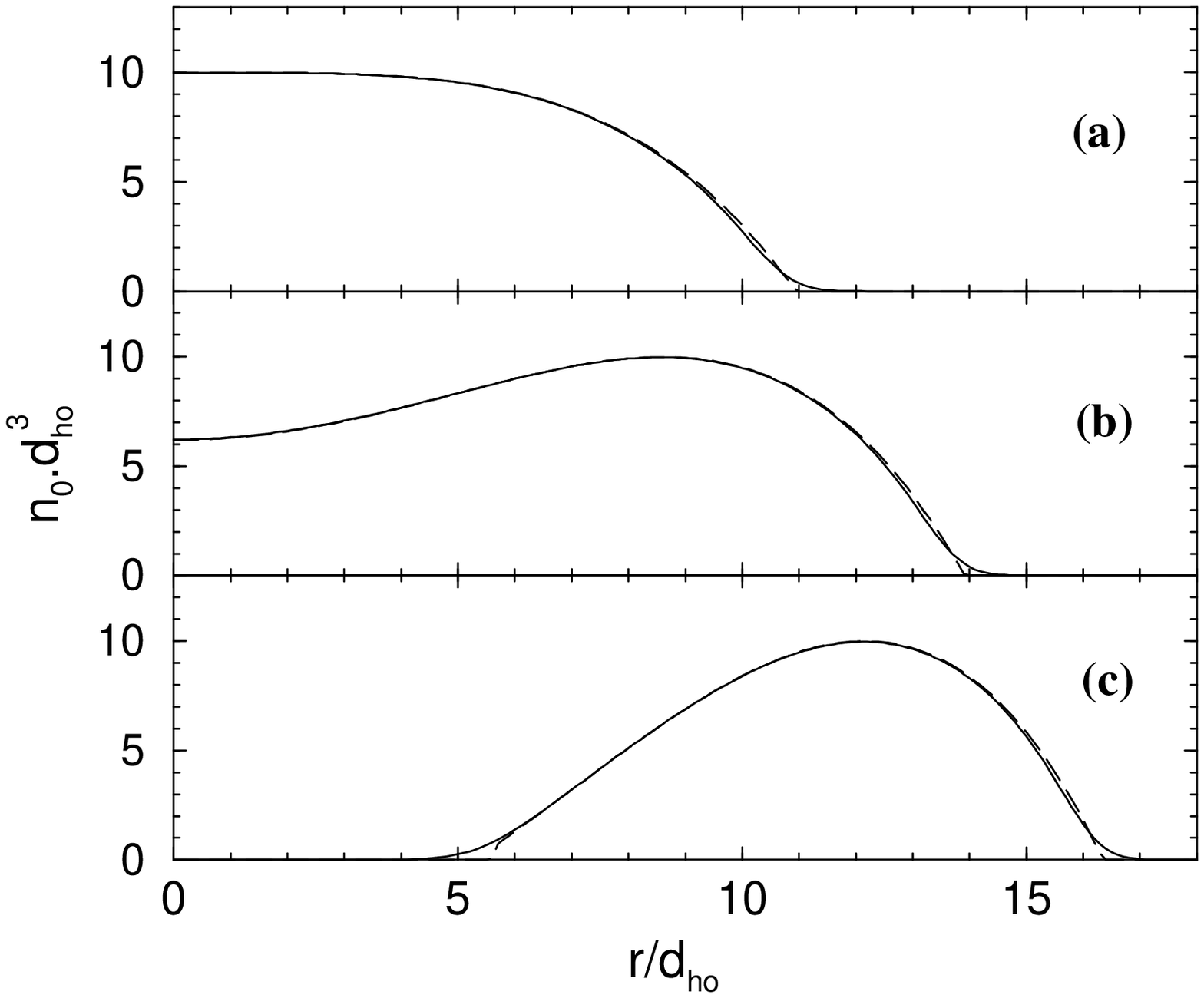}}
\caption{Radial profile of the condensate density:
Thomas-Fermi approximation results (dashed line) and the solution of 
the full differential equation of slave-boson approach (solid line) with 
$\beta'=1$ for the case without vortex ($m=0$). Thomas-Fermi results were previously 
published in ref. \cite{SBmanuscript}.
(a) Chemical potential $\mu'=0$, (b) $\mu'=1$, (c) $\mu'=2$ (units of 
$J$). The atomic density $n_0$ is 
given by equation (\ref{number}) . 
$r$ is the distance from the center of the trap and $d_{ho}$ is the 
harmonic oscillator length. Labels are in dimensionless units, like in all the followig plots.
(It should be noticed that the radial profile of the total density of particles is
monotonous with its maximum at $r=0$ \cite{shukla1}.)}
\label{fig1}
\end{figure}

In Fig.\ref{fig1} we show the condensate density profile of a trapped Bose gas at rest.
As the chemical
potential increases from (a) to (c), 
the total density of the Bose gas increases. It should be kept in mind that
the total density is always monotonically decreasing from the center of the trap (cf. Ref. \cite{shukla1}).
Due to increased interaction at higher total densities, depletation of the condensate reduces the 
condensate density at the center of the trap. As we can see in Fig.\ref{fig1}(a), at small chemical potential
(i.e. at low total density)
the condensate density is monotonically decreasing from the center, but at higher values of the chemical potential
depletion suppresses the condensate. This effect is stronger in regions with higher total densities (i.e. near the
center of the trap), such that at sufficiently high total densities the condensate is entirely destroyed
at the center (cf. Fig.\ref{fig1}(c)).

We also compare the full solution of Eq. (\ref{sbeq_novortex}) and the TFA solution of 
Ref. \cite{SBmanuscript} in Fig.\ref{fig1}. Although the difference between these solutions is small,
the full solution shows that the effect of the kinetic 
term is to smooth the abrupt changes of the TFA. Otherwise, the TFA is a very good approximation in 
the regime considered here. Moreover, the depletion of the condensate density at the center
shows a perfect agreement between the full solution and the TFA in this region.

\begin{figure}[ht] 
\centering 
\scalebox{0.5}{
\includegraphics{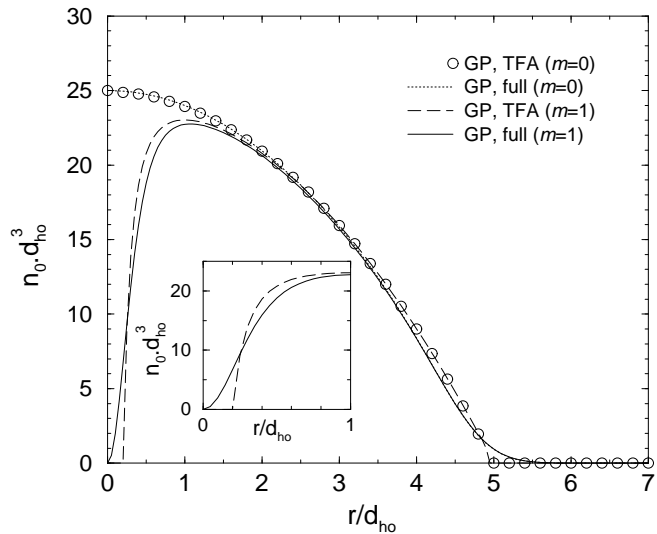}} 
\caption{Condensate density profiles obtained from the Gross-Pitaesvkii (GP) 
equation using the Thomas-Fermi approximation (TFA) and full numerical solution for 
the cases without vortex $(m=0)$ and with vortex 
$(m=1)$. The parameters are the chemical potential $\mu_{GP}=0.5$ and the total number of atoms
$N=9095$. $r$ is the distance from the 
center of the trap and $d_{ho}$ is the harmonic oscillator length. } 
\label{fig2} 
\end{figure}

In order to compare the slave-boson (SB) approach with resuls for the dilute Bose gas in the 
existing literature, we first evaluate the condensate density profile from the 
Gross-Pitaevskii (GP) equation (\ref{conventionalGP}) without $(m=0)$ and with $(m=1)$ vortex for the
TFA and for the full kinetic GP equation. The results are shown in Fig.\ref{fig2}. Without a vortex we observe the 
smoothing of the kinetic term near the discontinuity around $r=5$. In the vortex 
case, on the other hand, we find the formation of a core in the TFA. This core is also smoothed by 
the kinetic term in the full solution.

\begin{figure}[ht]
\centering
\scalebox{0.5}{
\includegraphics{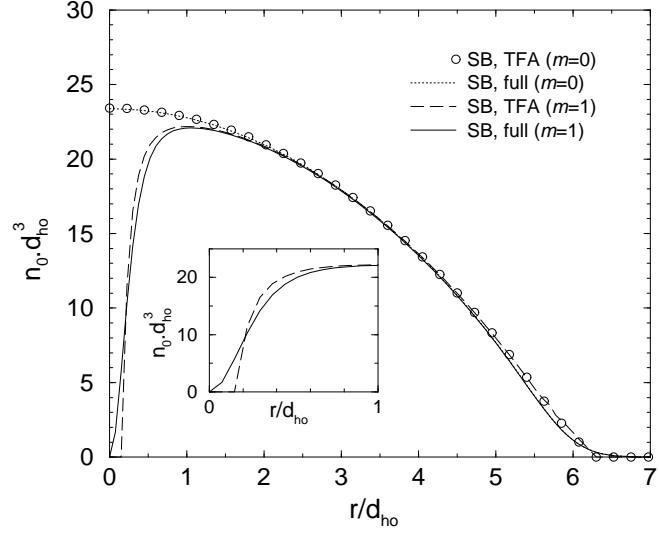}}
\caption{
Condensate density profiles obtained from slave-boson equation using 
Thomas-Fermi approximation (TFA) and full
numerical solution without vortex $(m=0)$ and with vortex $(m=1)$.
The parameters are $\beta'=10$, chemical potential
$\mu=-4.85 J$, and number of condensed particles $N_0=9112$. 
}
\label{fig3}
\end{figure}

Next, we evaluate the condensate density profile from the slave-boson approach, using Eq. (\ref{SBoriginal}),
for corresponding parameter values. The results are presented in Fig.\ref{fig3}. 
The parameters where chosen so that the number of condensed atoms roughly agree in Figs.\ref{fig2} and \ref{fig3}.

These results show a general similarity of the GP and SB profiles. The cases of the GP and the 
SB calculations without 
vortex indicate that there is a noticeable depletion of the central condensate density in the SB approach.
There is practically no difference between the TFA and the full solution. In the vortex case, the SB calculation 
shows that the core is also smooth when the kinetic term is added, as shown in the inset. 
Therefore the depletion in the center is overestimated by the TFA. 
This indicates that the formation of a vortex is less favorable than predicted by the TFA. 
Comparing the SB and the GP calculation in the vortex case, we also observe 
that the maximum of the SB profile is lower than that of the GP equation. Thus the depletion is more effective in
the SB approach than in the GP equation. 
These results are in agreement with the general idea that the SB approach takes into account 
higher order interaction terms, leading to a more pronounced depletion at the center.

\begin{figure}[ht]
\centering
\scalebox{0.5}{
\includegraphics[angle=0]{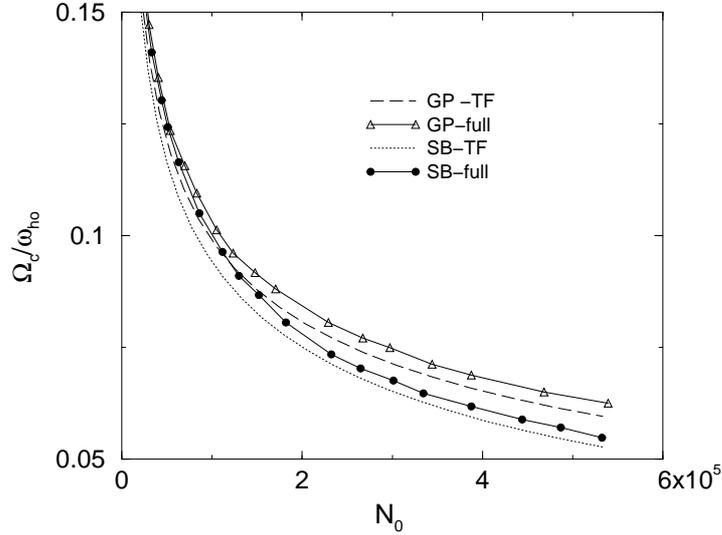}}
\caption{Critical angular velocity calculated in
Thomas-Fermi approximation and full calculations with kinetic
term for the Gross-Pitaevskii (GP) and Slave-Boson (SB) equations. $N_0$ 
is the 
number of condensed atoms. For the SB equation we used $\beta'=10$. 
} 
\label{fig4}
\end{figure}

The depletion at the center of the condensate should be observed experimentally by measuring the 
critical angular velocity of a stable vortex formation. In order to include strong interaction, the measurement
should be performed near a Feshbach resonance, where the scattering length is very large.
The critical angular velocity  can be evaluated in the SB approach by following 
the procedure outlined previously in Ref. \cite{Dalfovo2}. A vortex in a rotating condensate is stable if the 
action, given in Eq. (\ref{Scontinuous}), is smaller than the action of the condensate without vortex 
\begin{equation} 
S^{m=1}(\Omega) < S^{m=0} \; . 
\end{equation} 
Solving numerically Eq. (\ref{sbeq_vortex}) and using this stability criterion, we find the critical angular 
velocity $\Omega_{c}$ for a vortex formation. In Fig. \ref{fig4} we compile these results by plotting 
the critical angular velocity as a function of the number of condensed atoms. The TFA shows a lower critical 
angular velocity, both in the SB and in the GP approach, 
than the full numerical solutions. This is in qualitative agreement 
with previous results for the GP equation \cite{Sinha}. It can be simply understood in terms of our analysis 
above: Adding the kinetic term smooths the core generated by TFA. Thus the kinetic term
adds condensed particles to the center. This effect suppresses the vortex formation. 
A comparison indicates that for a small number of atoms, i.e. low 
density, the SB prediction is practically the same as that of the GP equation. However, as we increase the 
number of atoms, the SB 
approach significantly deviates from the GP equation, showing lower critical angular velocity for a vortex formation. 
This can be 
explained with the profile analysis described above: the SB approach causes a stronger depletion than the GP equation.
This difference is even enhanced as the total density increases. As a consequence, a stronger depletion will 
favor a vortex formation and thus to a lower critical angular velocity.

\section{Conclusion}

We employed the relaxation method to solve the slave-boson model. With this method we obtained results beyond the 
Thomas-Fermi approximation by including the kinetic term of the field equation. We noticed that the depletion 
of the condensate is overestimated by the Thomas-Fermi approximation. Therefore, in this approximation the critical 
angular velocities are smaller than in the full solution. We also compared the full slave-boson model with the full 
Gross-Pitaevskii equation in the limit of low temperatures. In this limit, the slave-boson condensate profiles are very 
similar to the Gross-Pitaevskii condensate profiles at low densities. As the density is increased, there is a significant 
enhancement of the depletion in the slave-boson approach as compared to the Gross-Pitaevskii equation. This 
causes the critical angular velocity of vortex formation to be lower than expected from the Gross-Pitaevskii 
equation. The latter results corroborate previous estimates within the Thomas-Fermi approximation done in 
Ref.\cite{SBmanuscript}. All these effects should be observable in a Bose gas near a Feshbach resonance,
where the scattering length is comparable with the mean distance of atoms.

\section*{Acknowledgements}

This work was supported by Coordenadoria de Aperfei\c{c}oamento de Pessoal de N\'ivel Superior 
(CAPES) and Deutscher Akademischer Auslandsdienst (DAAD).
A.G. also thanks Funda\c{c}\~ao de Amparo a Pesquisa do Estado de S\~ao Paulo (FAPESP) 
and Conselho Nacional de Desenvolvimento Cient\'ifico e Tecnol\'ogico (CNPq).


\vskip 1.0cm

\begin{thebibliography}{99}
\bibitem{note1} The functional-integral is not unique because of parameter-dependent transformation of fields.
\end{thebibliography}
\end{document}